\documentclass[aps,prl,showpacs,twocolumn]{revtex4}
\usepackage{bm}
\usepackage{epsfig}
\begin{document}

\title{Blackbody-radiation--assisted molecular laser cooling}
\author{I. S. Vogelius}
\affiliation{Department of Physics and Astronomy,
University of  Aarhus, 8000 {\AA}rhus C, Denmark}
\author{L. B. Madsen}
\affiliation{Department of Physics and Astronomy,
University of  Aarhus, 8000 {\AA}rhus C, Denmark}
\author{M. Drewsen}
\affiliation{QUANTOP - Danish National Research Foundation Centre for
  Quantum Optics, University of Aarhus,
8000 {\AA}rhus C, Denmark}

\begin{abstract}
The translational motion of molecular ions can be effectively
cooled sympathetically to temperatures below 100 mK in ion traps
through Coulomb interactions with laser-cooled atomic ions. The
distribution of internal rovibrational states, however, gets in 
thermal equilibrium with the typically much higher
temperature of the environment within tens of seconds. We
consider a concept for rotational cooling of such internally hot, but 
translationally cold heteronuclear diatomic molecular ions. The scheme
relies on a combination of optical pumping from a few specific
rotational levels into a ``dark state'' with redistribution of
rotational populations mediated by blackbody radiation.

\end{abstract}
\maketitle

Powerful techniques for manipulating, cooling and trapping atoms
have recently paved the way for extremely detailed 
investigations of atomic physics phenomena, as well as been essential
ingredients in the development of research fields such as
atom optics \cite{atom-optics}, physics of
trapped condensed dilute gases \cite{bec} and quantum information
\cite{qinfo}. Similar 
powerful techniques have not yet been established for molecules, but
such developments are expected to be equally rewarding in the
future.

Although molecules are routinely cooled internally and translationally
in supersonic expanded beams \cite{Miller88}, their high average
translational velocities ($\sim$ 100 - 1000 m/s) 
limit this method to experiments requiring only short
interaction times. The application of various laser cooling schemes,
already developed for atoms, is unfortunately hampered by the
radiative coupling of the many internal molecular energy levels which
impedes multiple laser-induced excitations. Recently, three very
different methods for trapping and cooling neutral molecules have,
however, been demonstrated. One of the schemes relies on optical
trapping of cold molecules produces by photoassociation of
laser-cooled atoms \cite{Takekoshi98,Fioretti98,Nikilov99,Nikilov00,Tolra01}. 
This approach is expected to work for homo- or heteronuclear dimers of
atoms amenable for laser cooling. Though the molecules are
produced translationally very cold (down to $\sim 100 \mu$K), the
population of the internal rovibrational states are so-far typically
spread over many levels. Another scheme is based on buffer gas cooling
of paramagnetic molecules held in a magnetic trap. In such experiments
all the molecular degrees of freedom have been cooled to $\sim 400$ mK
by collisions with a He buffer gas maintained at cryogenic
temperatures \cite{Weinstein98,Egorov01}. This method leaves the
molecules essentially in the ground state internally. Finally, beams of
neutral polar molecules have been decelerated by 
electrostatic fields \cite{Bethlem99}, and confined in an
electrostatic quadrupole trap \cite{Bethlem00} or a tabletop ``storage
ring'' \cite{Crompvoets01} at translational temperatures down to $\sim
10$ mK.

Molecular ions constitute another class of molecules that are
equally interesting to cool and manipulate. For decades
resonance-enhanced multiphoton ionization (REMPI) processes have been
exploited to produce state-specific molecular ions in beam experiments
\cite{rempi-beam}. The same procedure has also been applied in
connection with traps \cite{rempi-trap}, allowing for dramatically longer
interaction times for succeeding experiments.  
A more general technique for cooling to the internal ground state of
the molecular ions, is buffer gas cooling by He atoms in a cryogenic
environment, which has proven to cool trapped ionic molecules down to
10 K \cite{Gerl95}.  
Most recently, it has been demonstrated that  
molecular ions can be very effectively translationally cooled ($T <
100$ mK) into ion Coulomb crystals in a linear Paul trap through 
Coulomb interaction with laser cooled atomic ions
\cite{Moelhave00}. In such experiments, the frequency of the ions
external vibrational motion is determined by the trap potential, and
is typically 
$\sim 100-1000$ kHz. Rotational and vibrational frequencies of the
molecular ions, on 
the other hand, are many orders of magnitude larger ($\sim 10^{11}$--$10^{14}$
Hz) and, as a consequence, exchange of energy quanta between internal and
external degrees of freedom is forbidden by energy
conservation. Accordingly,  
the internal degrees of freedom are not sympathetically
cooled, and the rovibrational 
temperature of the heteronuclear molecules will reach 
equilibrium with the temperature of the surrounding trap
setup within tens of seconds\cite{Hechtfischer98,Amitay98}. The long
trapping times ($\sim$ 
hours) of these spatially localized ions, combined with long intervals
between collisions ($\sim$ minutes, as estimated by Langevin theory
\cite{Wineland98}),  however, opens for internal
cooling schemes which {\it rely on} and are not hampered by
blackbody radiation, and which are efficient on the timescale of
seconds.

\begin{figure}[htb]
\begin{center}
\epsfysize=5cm
\epsfxsize=6cm
\epsfbox{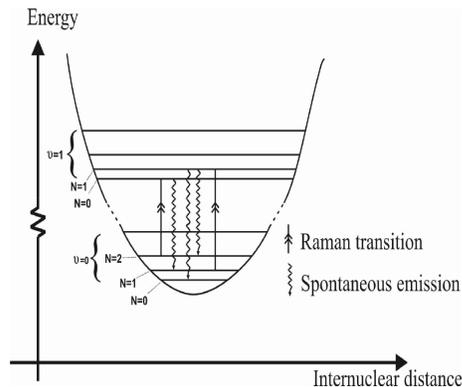}
\end{center}
\caption{\label{fig:raman-scheme} Figure showing the essential
  transitions needed for the Raman cooling scheme. A detailed
  explanation of the cooling principle is given in the
  text.} 
\end{figure}

In this Letter, we consider a cooling concept for the internal degrees
of freedom of trapped and translationally cold heteronuclear diatomic
molecular ions. Since at room temperature the vibrational degree of
freedom is frozen out for all the lighter species, i.e., the 
vibrational quantum number is equal to $\nu=0$, 
we focus on rotational cooling. Our concept involves the following processes:  
Pumping of population from ``pump states'' ($\nu=0, N=1$), and
($\nu=0,N=2$), where $N$ denotes the rotational quantum number, into
specific excited rovibrational states from which subsequent
spontaneous emission brings population back into either one of the
``pump states'', or into the ground state ($\nu=0, N=0$).
The latter is referred to as being a ``dark state'', since it is not
effected by the pumping fields. 
Finally, blackbody radiation (BR) is responsible for the feeding of the ``pump
states'' with populations from states with $N > 2$. In Fig.\
\ref{fig:raman-scheme}, a
sketch of the cooling concept, realized by utilizing two resonant,
dipole allowed Raman transitions ($\Delta N=0,\pm 2$), is
presented. In the absence of BR any initial
population in rotational states with $N=1$, and  $N=2$ would within a certain
time be optically pumped into the ``dark state'' ($\nu=0, N=0$) due to
the selection rules $\Delta N=\pm 1$ for spontaneous emission. In the
presence of BR effective cooling into the internal
ground state is possible from an initial thermal
distribution with significant population in states with $N>2$, as
long as the rate of optical pumping and the rate of spontaneous
emission from the vibrationally excited state are higher than the
redistribution rate among the rotational states due to BR.  
The cooling time will in such cases be set by the inverse of
the typical rotational redistribution rate. 

In an experiment the Raman transitions will be driven by
pulsed lasers. In that case, high population transfer is assured if
each Raman pulse saturates its transition. This means the pulse 
time, $\tau$, times the Raman coupling frequency,
$\Omega_{\mbox{\scriptsize{R}}}$, should fulfill
(i) $\tau \Omega_{\mbox{\scriptsize{R}}} \agt 10$, where
$\Omega_{\mbox{\scriptsize{R}}} \simeq \Omega^2/\delta$ with $\Omega$
being the typical Rabi frequency of the dipole allowed electronic
transitions and $\delta$ the corresponding detuning. 
Furthermore, a small incoherent scattering rate is required, leading to (ii)
$\tau \Gamma_{\mbox{\scriptsize{scat}}}
\alt 0.01$, with
$\Gamma_{\mbox{\scriptsize{scat}}}\simeq \Omega_{\mbox{\scriptsize{R}}} A
/ \delta$, $A$ being the pertaining Einstein
coefficient. 

\begin{figure}[htb]
\begin{center}
\epsfysize=5cm
\epsfbox{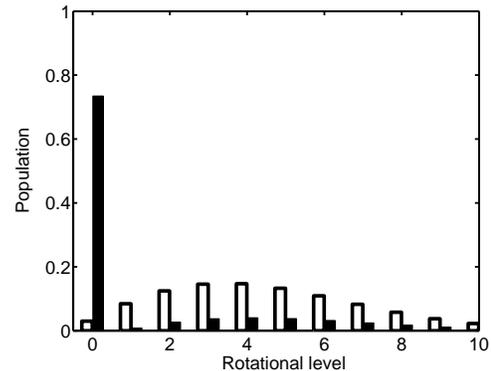}
\end{center}
\caption{\label{fig:MgHplus} Cooling of MgH$^+$ by the Raman scheme
  (see Fig.\ \ref{fig:raman-scheme}). 
  The unfilled columns
  represent the initial rotational populations corresponding to a
  thermal distribution at $T=300$ K. The filled columns represent the
  population distribution after 100 s of cooling. See the text for
  details on the cooling parameters.} 
\end{figure}

We have modelled the dynamics of the BR--assisted cooling by
rate equations including change in population of 
the involved rovibrational states due to both 
spontaneous and stimulated processes.
In practice this means that the calculations require the knowledge of
Einstein $A$ and $B$ coefficients for a selection of rovibrational 
transitions. In  short, the dipole moment functions and potential curves 
are obtained {\it ab initio} 
using Gaussian94 \cite{gaussian94}. The energies of the levels of
relevance are in
agreement with published data within 1.5\% \cite{Huber79}. 
Einstein coefficients are subsequently calculated from the dipole
moment function and potential energy curve using the Numerov
method\cite{leroy}. 
Details of the calculations will be presented elsewhere
\cite{Vogelius-tocome}. 
The initial populations are taken to be Boltzmann distributed at room
temperature ($T$ = 300 K).  
In Fig.\ \ref{fig:MgHplus}, we present the results of such simulations
in the case of MgH$^+$ which has been cooled translationally in the
laboratory\cite{Moelhave00}.  
The unfilled columns represent the initial rotational
populations, while the black columns show the populations after 100
seconds of cooling. In the simulation, Raman
pulses couple the vibrational states of the X$^1\Sigma^+$ potential curve  
via the electronically excited A$^1\Sigma^+$ state, using lasers in the
wavelength range $\sim 279$ nm \cite{Huber79}. 
We find that the saturation and scattering conditions (i) and (ii)
above can be fulfilled for the pumped transitions from any rotational  
sub-state for 10 ns pulses with intensities  $\sim$ 100 kW/cm$^2$. 
The repetition rate of the Raman pulses was 100 Hz, which is much
higher than the typical rate for 
rotational transitions due to BR.
The figure shows that more than 70\% of the population can be 
accumulated in the rovibrational ground state, equivalent to a
thermal distribution at 8.5 K. 

\begin{figure}[htb]
\begin{center}
\epsfysize=5cm
\epsfbox{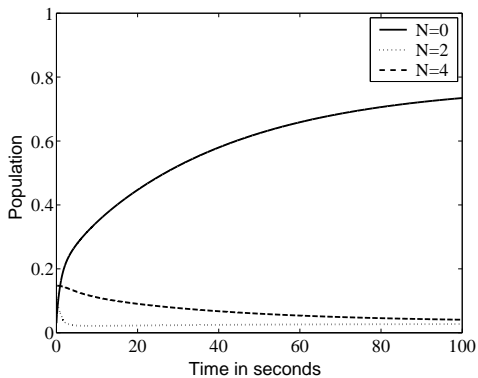}
\end{center}
\caption{\label{fig:timescale} Evolution of the populations of the
  rotational levels with quantum numbers $N=0,2,4$ as a function of
  cooling time for MgH$^+$.} 
\end{figure}

Figure \ref{fig:timescale} shows the evolution of the population in
three representative rotational levels of MgH$^+$ for the Raman scheme
discussed above. After 100 s of
cooling the system has practically reached steady--state. 

\begin{figure}[htb]
\begin{center}
\epsfysize=5cm
\epsfbox{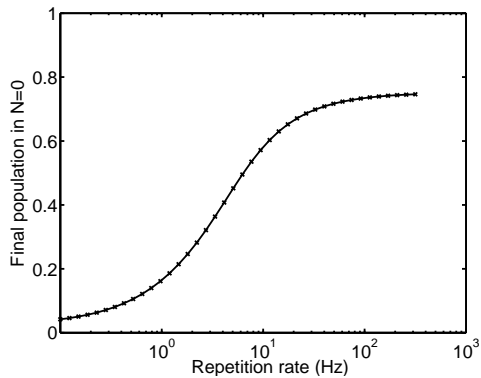}
\end{center}
\caption{\label{fig:rep-rate} Population in the rovibrational ground
  state of MgH$^+$ after 100 s of cooling as a function of the repetition rate
  of the pumping lasers. Each laser pulse saturates the transition
 it pumps.}
\end{figure}

Figure \ref{fig:rep-rate} explores the sensitivity of the cooling
efficiency for MgH$^+$ on the repetition rate of the Raman pulses for
a fixed cooling time of 100 s. 
Already for modest repetition rates, say 20--30 Hz, we observe a
significant cooling efficiency and repetition rates larger than 100
Hz do not significantly increases the cooling efficiency. This is
experimentally very encouraging since laser systems with nanosecond
pulses typically have repetition rates in the interval 10 -- 100 Hz. 

The Raman scheme discussed above is effective for molecular ions which
have an excited electronic state that can be addressed by laser light
in the visual or near-visual range. If such excited states are absent,
a scheme based on continuous 
wave near infra-red (IR) sources introducing direct couplings {\it within}
the ground electronic potential curve can be applied. This direct
scheme works as follows:  
One IR source couples ($\nu=0,N=1$)--($\nu=2,N=0$) which, although dipole
forbidden, for typical molecules is easily 
saturated due to the anharmonicity of the potential curve.
The excited ($\nu=2,N=0$) cascades down by dipole allowed
transitions, first to the ($\nu=1,N=1$)
excited state and then to the ($\nu=0,N=0$) dark state or the 
($\nu=0,N=2$) state. The latter state is pumped by another laser into
the ($\nu=1, 
N=1$) state, from which decay into the dark state or back into the
($\nu=0,N=2$)  
state will take place.
   
\begin{figure}[htb]
\begin{center}
\epsfysize=5cm
\epsfbox{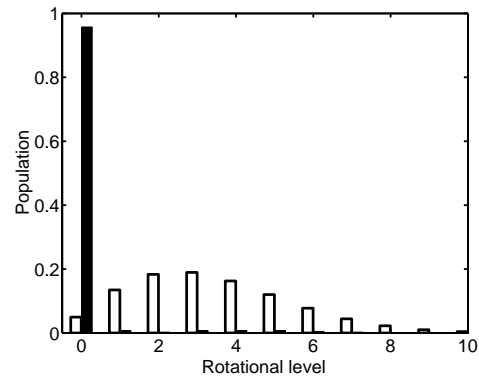}
\end{center}
\caption{\label{fig:ArHplus} Cooling of ArH$^+$ by the direct
  scheme. The unfilled columns
  represent the initial rotational populations corresponding to a
  thermal distribution at $T=300$ K. The filled columns represent the
  population distribution after 50 s of cooling. See the text for
  details on the scheme, and 
  the cooling parameters.} 
\end{figure}

As an example of the application of this direct scheme we consider
ArH$^+$ which is supposed to have no stable excited electronic
states \cite{Schutte01a}. Fig.\ 
\ref{fig:ArHplus} shows the initial thermal ($T=300$ K) 
distribution over rotational states for ArH$^+$ and the final 
steady--state distribution after 50 s of
cooling. In the cooled distribution, more than 
95\% of the population is in the X$^1\Sigma^+$ rovibrational ground
state, equivalent to a temperature of 7 K. ArH$^+$ is a  
strong vibrational infrared emitter \cite{Picque00,Rosmus79}, and 
the rotational transition rates are of the same order of magnitude as
in MgH$^+$\cite{Laughlin87} and,
consequently, the cooling is very effective. The IR
stimulated processes which drives the ($\nu=0,N=1$)--($\nu=2,N=0$) and
($\nu=0,N=0$)--($\nu=1,N=1$) transitions, requires lasers with
wavelengths around 
$1.9 \mu$m and $3.8 \mu$m, respectively \cite{Schutte}. 
These wavelengths are conveniently covered by near-IR 
continuous-wave optical parametric oscillators (OPO's) \cite{Kovalchuk01}. To
ensure that the scheme remains insensitive to the intrinsic linewidth
and drifts in the OPO, one should require an effective rate for the
pumped transitions of $\sim 90$\% of the $A$-coefficients, 
at a detuning of $\sim 10$ MHz. This is fulfilled at an
intensity of a few hundreds of W/cm$^2$, which is realistic using a
laser beam with a few tens of mW power focused to a beam waist of
$\sim 100 \mu$m. Note that such a waist is much larger than the
localization of translationally cold ions \cite{Moelhave00}.   

To make sure that all magnetic sublevels of the rotational states are
addressed by the pump laser fields, the polarization of the
pumping radiation must be varied on the time scale of the optical
pulses in the case of the Raman scheme and at a rate faster than
rotational transition by BR is the case of the direct scheme. In both
cases this can be done by using a modulated Pockels cell eventually
combined with splitting the pulses to enter the trap region from
different directions.  

Here, we have focused on the description of schemes for cooling the
rotational degree of freedom of molecular ions with the ground state symmetry, 
$^1\Sigma^+$ and without hyperfine splittings. Such splittings,
however, are typically much smaller, e.g. $\sim 1$ MHz in BeH$^+$
\cite{Fiser96} ,  than the bandwidth $\sim 1-30$ GHz of typical pulsed
laser systems applicable in the Raman scheme, and, hence, all
hyperfine levels will be addressed by the pump fields.
Also, the schemes can be extended to molecular ions with more
complicated ground state configurations at the cost of more detailed
considerations of the laser systems used \cite{Vogelius-tocome}. 

If one considers an implementation of the schemes in connection with
cooling in storage rings \cite{Hechtfischer98,Amitay98} a
larger focal spot of the laser would be required
in order to obtain overlap with the ion beam. 
This significantly increases the laser power requirements, 
and leads us to conclude that the pulsed Raman scheme could be
considered while the direct scheme would be impossible, due to the limited
power of present day OPO systems.

The state-selected and strongly localized molecular ions produced by
the above presented cooling schemes have many potential applications
and will be interesting for 
a large variety of studies including controlled chemical reactions,
implementations of quantum logics, 
and for the mimicking of conditions in the interstellar medium 
where many small and cold molecular ions play an important role \cite{chem}.

In conclusion, we have shown how initially
translationally cold, trapped molecular ions can be internally cooled by simple
optical pumping schemes when assisted by the blackbody radiation.
The schemes are simple, and robust, and the light source requirements
are modest compared to state--of--the--art laser systems.

\bigskip
\noindent
L.B.M. is supported by the Danish Natural Science Research Council
(Grant No.\ 51-00-0569). 
M.D. is supported by the Danish National Research Foundation through
the Quantum Optics Centre QUANTOP as well as by the Carlsberg Foundation.

\end{document}